\documentclass{svmult}

\usepackage{makeidx}         
\usepackage{graphicx}        
\usepackage{multicol}        
\usepackage[bottom]{footmisc}

\usepackage{amsmath,amssymb,mathrsfs,eucal}

\makeindex             

\begin{document}
\newcommand{\ip}[2]{{\langle #1\,|\, #2\rangle}}
\newcommand{\RR}{{\mathbb R}}
\newcommand{\HH}{{\mathscr H}}
\newcommand{\KK}{{\mathscr K}}
\newcommand{\LL}{{\mathscr L}}
\newcommand{\FF}{{\mathscr F}}
\newcommand{\PP}{{\mathscr P}}
\newcommand{\Sch}{{\mathscr S}}
\newcommand{\cA}{{\mathcal A}}
\newcommand{\cR}{{\mathcal R}}
\newcommand{\cV}{{\mathcal V}}

\newcommand{\OO}{{\mathcal O}}
\newcommand{\sN}{{\mathscr N}}

\newcommand{\QQ}{{\mathcal Q}}
\newcommand{\WW}{{\mathscr W}}

\newcommand{\VV}{{\cal V}}

\newcommand{\Af}{{\mathcal{A}}}
\newcommand{\Ff}{{\mathcal{F}}}
\newcommand{\Of}{{\mathcal{O}}}
\newcommand{\Sf}{{\mathcal{S}}}
\newcommand{\Uf}{{\mathcal{U}}}
\newcommand{\Vf}{{\mathcal{V}}}

\newcommand{\ux}{\underline{x}}

\newcommand{\Had}{{\sf Had}}
\newcommand{\WF}{\mathop{\rm WF}}
\newcommand{\Tr}{\mathop{\rm Tr}}
\newcommand{\II}{{\boldsymbol{1}}}
\newcommand{\CoinX}[1]{C_0^\infty({#1})}

\title*{Quantum Energy Inequalities and Stability Conditions in Quantum
Field Theory\thanks{Talk given at the symposium `Rigorous Quantum Field
Theory' held in honour of the 70th
birthday of Jacques Bros, Paris,
July 2004.}}
\titlerunning{QEIs and stability conditions in QFT}
\author{Christopher J Fewster}
\authorrunning{CJ Fewster}
\institute{Department of Mathematics, University of York, Heslington,
York YO10 5DD, United Kingdom.
\texttt{cjf3@york.ac.uk}}
%
%
\maketitle
\begin{abstract}
We give a brief review of quantum energy inequalities (QEIs) and then discuss
two lines of work which suggest that QEIs are closely related
to various natural properties of quantum field theory which may all be
regarded as stability conditions.
The first is based on joint work with Verch, and draws connections
between microscopic stability (microlocal spectrum
condition), mesoscopic stability (QEIs) and macroscopic
stability (passivity). The second direction considers QEIs for a
countable number of massive scalar fields, and links the existence and scaling
properties of QEIs to the spectrum of masses. The upshot is 
that the existence of a suitable QEI with polynomial scaling
is a sufficient condition for the model to satisfy the
Buchholz--Wichmann nuclearity criterion. We briefly discuss 
on-going work with Ojima and Porrmann which seeks to gain a deeper
understanding of this relationship. 
\end{abstract}
 
\section{Introduction}
\label{sec:QEIs}

The stress-energy tensor $T_{ab}$ of the real scalar field, in common with
those corresponding to most models of classical
matter,\footnote{The main exception is the nonminimally coupled scalar
field, satisfying the field equation $(\Box+m^2+\xi R)\phi=0$ with
$\xi\not=0$.}
obeys the dominant energy condition (DEC): for any
future-directed timelike vector $u^a$, the contraction
$T^{a}_{\phantom{a}b}u^b$ is itself timelike and future-directed. This
may also be stated as the inequality $T_{ab}u^a v^b\ge 0$ for all pairs
of future-directed timelike vectors $u^a$ and $v^a$. In the special case
$v^a=u^a$, we recover the weak energy condition (WEC), $T_{ab}u^au^b\ge
0$, i.e., the energy density is nonnegative according to any observer.

In classical general relativity, energy conditions of this type play a key
role, guaranteeing the stability of gravitational collapse
(singularity theorems~\cite{HawkingEllis}), the stability of Minkowski space as a `ground
state' of the theory (positive mass theorems~\cite{Witten81,LudVic}) and also excluding certain
exotic causal structures (see e.g., Hawking's discussion of chronology protection~\cite{HawkingCPT}). 
However, as has been known for a long time~\cite{EGJ}, the WEC (and
hence DEC) are violated in quantum field theory. It is easy to give a simple proof in
Minkowski space: suppose that $\varrho=T_{ab}u^a u^b$, where $u^a$ is now a smooth
timelike vector field, and let $f$ be a nonnegative smooth function of
compact support. We need assume only that the smeared field $\varrho(f)$
is (essentially) self-adjoint on the Hilbert space of the theory and that
there is a vacuum state $\Omega$ in the operator domain of $\varrho(f)$ such that 
$\ip{\Omega}{\varrho(f)\Omega}=0$ but which is not annihilated by
$\varrho(f)$, i.e., $\varrho(f)\Omega\not=0$. Writing the
spectral measure of $\varrho(f)$ as $\D E(\lambda)$, these last two
properties tell us that the probability measure 
$\ip{\Omega}{\D E(\lambda)\Omega}$ on $\RR$ has zero expectation, but that its
support is not simply the set $\{0\}$. Accordingly $\varrho(f)$ must have
some negative spectrum.\footnote{The same conclusion is easily drawn by
examining the expectation values of $\Omega+\lambda\varrho(f)\Omega$ for
small $\lambda$.} Clearly the same argument applies in many
circumstances, and for observables other than energy density.

Thus, the classical pointwise energy conditions are simply incompatible with the structures
of quantum field theory. Further analysis of particular models shows
that the pointwise energy density is typically unbounded from below as a
function of the quantum state, and this can be proved for all theories
with a suitable scaling limit~\cite{Lisbon}.

This fact raises questions concerning the applicability of the
singularity, positive mass and chronology protection results where
quantised matter is concerned. Many authors have also sought to
exploit quantum fields to support metrics (including wormhole or warp
drive models) which require WEC-violating matter distributions. It is
therefore important to understand whether the classical energy conditions are
irretrievably lost, or whether one can identify some remnant 
in the quantum theory. This contribution will discuss a promising
candidate: a group of results known as Quantum Energy Inequalities (QEIs), and will in
particular focus on their emerging connections with other well-known stability
conditions in quantum field theory, namely the microlocal spectrum
condition, passivity and nuclearity. The hope is that, by unravelling
these connections, further insight is provided into the nature of
quantised matter and its (gravitational) stability. 

It is a particular pleasure to dedicate this contribution to Jacques Bros, in view
of his influential contributions to both microlocal analysis and the
description of thermal behaviour in quantum field theory.

\section{Quantum Energy Inequalities}

As mentioned above, the pointwise energy conditions are unavoidably
and severely violated in quantum field theory. However, observations at
individual spacetime points are not
physically achievable in any case (owing to the uncertainty principle),
so it is more natural to consider weighted averages of the stress-energy
tensor over a spacetime volume. 

\begin{definition} 
Let $\WW$ be a class of second-rank tensors on spacetime, and
$\Sf$ a class of states of the theory. If, for each $\tens{f}\in\WW$,
the averaged
expectation values $\int \D{\rm vol}(x) \langle  T_{ab}(x)\rangle_\omega f^{ab}(x)$ are bounded from below
as $\omega$ runs over $\Sf$, we say that the theory obeys a \emph{Quantum Energy
Inequality} (QEI) with respect to $\WW$ and $\Sf$.
\end{definition}
One generally aims to find an explicit lower bound $-\QQ[\tens{f}]$ so that the QEI can be written as an inequality
\begin{equation*}
\int \D{\rm vol}(x) \langle T_{ab}(x)\rangle_\omega f^{ab}(x) \ge -\QQ[\tens{f}] \qquad \forall \omega\in\Sf\;.
\end{equation*}
Where $\WW$ consists of tensors of a particular form e.g., $f^{ab}=u^a u^b$ or
$f^{ab}=u^a v^b$ for timelike vector fields $u^a$, $v^a$, we use more
specific terms, e.g., Quantum Weak Energy Inequality (QWEI) or Quantum
Dominated Energy Inequality (QDEI). Of course a similar approach could be
adopted other quantities of interest.

For the most part, QEIs have been developed for averages along timelike
curves, rather than over spacetime volumes, in which case the weights may be
thought of as being singularly supported on a curve. By threading a
spacetime volume by worldlines, these bounds imply 
the existence of spacetime-averaged QEIs, which may also be obtained
directly, as sketched below. It is known that compactly supported
weighted averages over spacelike
hypersurfaces~\cite{FHR} or null lines~\cite{FewsterRoman03} are not generally bounded from
below,  except for two-dimensional conformal fields~\cite{Flan,FewsterHollands}.

QEIs were first proposed by Ford~\cite{Ford78}, who realised that
suitable bounds of this type would be sufficient to prevent macroscopic
violations of the second law of thermodynamics arising from negative
energy phenomena in quantum field theory. They have since 
been established for the free
Klein--Gordon~\cite{Ford91,FordRoman95,FordRoman97,FPstat,FewsterEveson,FTi,AGWQI,Flan,Vollick2000,Flanagan02},
Dirac~\cite{Vollick2000,FVdirac,FewsterMistry}, 
Maxwell~\cite{FordRoman97,Pfenning_em,FewsterPfenning} and
Proca~\cite{FewsterPfenning} quantum fields in
both flat and curved spacetimes, the Rarita--Schwinger field in
Minkowski space~\cite{YuWu}, and also for general unitary
positive-energy conformal field theories in two-dimensional Minkowski space~\cite{FewsterHollands}. We will not give a full history of the development of
the subject, referring the reader to the recent reviews~\cite{Lisbon,Roman_review}.  
To give a flavour of the sort of results obtained, we give an example in
which the energy density of a scalar field of mass $m$ is averaged along the inertial
trajectory $(t,0)$ in Minkowski space. It can be obtained by elementary
means~\cite{FewsterEveson} or as a special case of the rigorous 
result~\cite{AGWQI}. Set $\rho=T_{ab}u^a u^b$, where
$\tens{u}=\partial/\partial t$. Then the QWEI
\begin{equation}
\int \D t\,\langle\varrho(t,0)\rangle_\psi |g(t)|^2 \ge-\QQ[g]:= -\frac{1}{16{\pi}^{3}}\int_0^\infty
\D u\,u^4 \vartheta(u-m) |\widehat{g}(u)|^2\;,
\label{eq:FEQI}
\end{equation}
holds for all Hadamard states $\psi$ (see below) and smooth compactly supported $g$. Here
$\widehat{g}$ denotes the Fourier transform\footnote{Our convention for the Fourier transform
is $\widehat{g}(u)=\int \D t\, \E^{\I ut}g(t)$ etc.} and $\vartheta$ is the Heaviside function. The bound is 
finite, owing to the rapid decay of $\widehat{g}$. In fact the bound
given in~\cite{FewsterEveson} is slightly tighter than this, but~\eqref{eq:FEQI}
will suffice for our present purposes. 

For later reference, let us
note the scaling behaviour of the bound~\eqref{eq:FEQI}. Replacing $g$ by
$g_\tau(t)=\tau^{-1/2}g(t/\tau)$, so that $\tau$ controls the `spread'
of the weight, one may show that
\begin{equation*}
\QQ[g_\tau] = \left\{ \begin{array}{cl} 
O(\tau^{-4}) & {\rm as}~\tau\to0^+\\
O(\tau^{-\infty}) & {\rm as}~\tau\to\infty\end{array}\right.
\end{equation*}
for $m>0$, where the notation $O(\tau^{-\infty})$ indicates faster-than-inverse-polynomial
decay. In the massless case, it turns out that $\QQ(g_\tau)\propto
\tau^{-4}$ for all $\tau>0$. We note that the $\tau\to 0^+$ limit, which
corresponds to sampling at a point, is consistent with the pointwise unboundedness
below of the energy density. For intermediate scales, the QWEI allows
for a limited violation of the classical WEC; bounds of this type
therefore appear to be the natural remnant of the WEC in quantum field
theory. 

We mention briefly that related bounds appear elsewhere in quantum
field theory~\cite{Marecki} and quantum mechanics~\cite{EFV}; 
QEIs have also been used to place constraints on exotic
spacetimes~\cite{FRworm,FPwarp,Roman_review}.

\section{Stability at Three Scales}

The work described in this section, conducted with
Verch~\cite{FVpassive} and building on earlier work~\cite{AGWQI,SV1},
uncovers a circle of connections between stability conditions operating
at three different scales: the microscopic (Hadamard
condition/microlocal spectrum condition), mesoscopic (QEIs) and
macroscopic (thermodynamic stability, expressed by the notion of
passivity~\cite{PW}). Each connection takes the form of a rigorous theorem; the
reader should be cautioned, however, that the conclusions and hypotheses
of successive links do not match perfectly. Moreover, two of the links (mesoscopic
to macroscopic, and macroscopic to microscopic) are obtained in greater
generality than the particular setting of quantum field theory on curved
spacetimes, while the microscopic to mesoscopic link is currently known
only for particular models of quantum field theory. Thus the existence of these connections should be
regarded as indicative of a close relationship between these three
stability conditions, rather than of proving their equivalence. In part,
this work gives a precise expression to Ford's original insight~\cite{Ford78}, that
bounds of QEI type would suffice to prevent macrosopic violations of the
second law of thermodynamics.

\subsection{Microscopic Stability: the Hadamard Condition}

Stability of quantum field theory at the microscopic scale is (partly) expressed
by the Hadamard condition, which requires that the singular structure of
the two-point function takes a form determined for nearby
points by the local geometry~\cite{KayWald} of spacetime. As first shown by
Radzikowski~\cite{Radzikowski96}, this may be reformulated as
a condition on the wave-front set~\cite{HormanderI} of the two-point function.\footnote{Appropriate
conditions on higher $n$-point functions were given in~\cite{BFK96}. For non-initiates: the wave-front set
$\WF(S)$ of a distribution $S$ on a manifold $M$ is a subset
of the cotangent bundle $T^*M$ which encodes the singular structure of
$S$. Singularities are classified in terms of the (lack of) decay of local
Fourier transforms of $S$ in different directions.} By passing
to a Hilbert space representation, however, one obtains a very simple
formulation of the Hadamard condition~\cite{SVW,FVpassive} (cf.
also~\cite{BrunettiFredenhagen}): a
state of the scalar field on $(M,\tens{g}$) is Hadamard if and only if 
it may be represented by a vector $\psi$ in some Hilbert space
representation of the theory so that $f\mapsto \Phi(f)\psi$ is a vector-valued distribution
whose wave-front set  obeys
\begin{equation}
\WF(\Phi(\cdot)\psi) \subset \VV^-\;,
\label{eq:WF}
\end{equation}
where $\Phi$ is the field and 
\begin{equation*}
\VV^- =\{(x,\tens{k})\in T^*M: g^{ab}k_a k_b\ge 0,~\tens{k}~\hbox{past directed}\}
\end{equation*}
is the bundle of past-pointing causal covectors (our signature
convention is $+---$). This has the following practical upshot. Suppose $f$ is smooth and
compactly supported within some coordinate patch, with coordinates
$x^\alpha$ so that $\partial/\partial x^0$ is future-pointing and
timelike. Let $V$ be any closed cone in $\RR^4$ consisting of $k$ such
that the covector field $k_\alpha \D x^\alpha$ is nowhere causal and past-directed on the
coordinate patch. In particular, $V$ could be the half-space $V=\{k\in\RR^4: k_0\ge
0\}$. Then 
\begin{equation*}
I(k) := \left\| \int \D^4 x\, \E^{\I k_\alpha x^\alpha} f(x) \Phi(x)\psi
\right\|
\end{equation*}
is of rapid decay in $V$; that is, it decays more rapidly than any inverse polynomial in
the Euclidean norm of $k$ as $k\to\infty$ in $V$.
Moreover, the same is true if $f$ is replaced
by a partial differential operator with smooth coefficients compactly
supported in the coordinate patch. 

Microlocal formulations of the Hadamard condition
are also known for the Dirac~\cite{Kratzert,Hollands01,SV2}, Maxwell and Proca
fields~\cite{FewsterPfenning}. They may be regarded as local remnants of
the spectrum condition, i.e., the Minkowski space requirement that the
joint spectrum of the generators $P_\mu$ of spacetime translations
should lie in the future causal cone.\footnote{That the forward cone
appears in the spectrum condition, but the backward cone in~\eqref{eq:WF}, is the result of an unfortunate clash of conventions.} 

\subsection{From Microscopic to Mesoscopic}

We now show how QEIs may be derived from the Hadamard condition, using an
argument based on that of~\cite{AGWQI}. The classical Klein--Gordon field $\phi$ obeying
$(\Box+m^2)\phi=0$ on spacetime $(M,\tens{g})$ has stress-energy tensor
\begin{equation*}
T_{ab} = \nabla_a\phi \nabla_b\phi -
\frac{1}{2}g_{ab}g^{cd}\nabla_c\phi\nabla_d\phi + \frac{1}{2}g_{ab}m^2\phi^2\;,
\end{equation*}
and obeys the WEC and DEC because the relevant
contractions of $T_{ab}$ can be decomposed as sums of squares. Let us
therefore consider -- as representing the most general classical energy
condition -- any tensor field $f^{ab}$ for which 
\begin{equation}
T_{ab}f^{ab} = \frac{1}{2}\sum_j (P_j\phi)^2\;,
\label{eq:Tfsqs}
\end{equation}
where the $P_j$ are finitely many linear partial differential operators (possibly of degree zero) 
with smooth real coefficients of compact support. Clearly $T_{ab}f^{ab}\ge
0$ for classical fields $\phi$. 
For simplicity, assume that the supports of the $P_j$ are contained
within a single coordinate patch of $(M,g)$ writing the coordinates as
$x^\alpha$ and assuming as above that $\partial/\partial x^0$ is future-pointing and timelike.
Write also $g(x)=|\det g_{\alpha\beta}(x)|$.

Let $\psi_0$ be a fixed Hadamard reference state. Defining the
stress-energy tensor ${:}T_{ab}{:}$ by point-splitting and normal ordering with respect
to $\psi_0$, we have
\begin{equation*}
\langle{:}T_{ab}(x){:}\rangle_\psi f^{ab}(x) =
\frac{1}{2\sqrt{g(x)}}F(x,x)
\end{equation*}
for any Hadamard state $\psi$, where
\begin{equation*}
F(x,y) = (g(x)g(y))^{1/4}\sum_j \left[
\left\langle (P_j\Phi)(x)(P_j\Phi)(y)\right\rangle_\psi - \langle (P_j\Phi)(x)(P_j\Phi)(y)\rangle_{\psi_0}\right]
\end{equation*}
is smooth (owing to the common singularity structure of Hadamard
two-point functions) and symmetric (because two-point functions have a
state-independent antisymmetric part). Thus we may write
\begin{eqnarray*}
\int \D{\rm vol}_{\tens{g}}(x) \langle{:}T_{ab}(x){:}\rangle_\psi f^{ab}(x) &=&
\frac{1}{2}\int \D^4x\,\D^4y\, F(x,y)\delta^{(4)}(x-y)\nonumber\\
&=&
\int_{k_0\ge 0} \frac{\D^4k}{(2\pi)^4} \int \D^4x\,\D^4y\, \E^{-\I k\cdot(x-y)}F(x,y)\;,
\end{eqnarray*}
where we have used the Fourier representation of the Dirac-$\delta$ and
symmetry of $F$ to restrict the outer domain of integration. Now the
inner integral is $A(k;\psi)-A(k;\psi_0)$, where 
\begin{equation*}
A(k;\psi) := \sum_j\left\| \int \D^4x\, \E^{\I k\cdot x} g(x)^{1/4}
P_j\Phi(x)\psi\right\|^2\ge 0\;,
\end{equation*}
and so we obtain the QEI 
\begin{equation}
\int \D{\rm vol}_{\tens{g}}(x)\, \langle {:}T_{ab}(x){:}\rangle_\psi f^{ab}(x)\ge
-\int_{k^0\ge 0}
\frac{\D^4k}{(2\pi)^4} A(k;\psi_0)\;, 
\label{eq:QEI1}
\end{equation}
the left-hand side of which depends on the reference state $\psi_0$, but not on $\psi$. 
The key point now is that the microlocal form of the Hadamard
condition entails that $A(k;\psi_0)$ is of rapid decay in the half-space $k_0\ge 0$. Thus the integral
on the right-hand side of~\eqref{eq:QEI1} exists and is finite.
We conclude that the real linear scalar field obeys a QEI with respect to
the class of weights delineated by~\eqref{eq:Tfsqs} and the class of
Hadamard states. The same argument would apply to a suitable class of
adiabatic states~\cite{JunkerSchrohe} in which one replaces the smooth wave-front set
by a wave-front set modulo Sobolev regularity. 

Note that this QEI applies to the normal ordered stress-energy tensor,
rather than the renormalised tensor.\footnote{To form the renormalised
tensor, we begin by splitting points as above, 
but then subtract appropriate derivatives of the locally determined Hadamard parametrix, rather
than the two-point function of a reference state.} By adding a term to
the both sides which depends on the renormalised stress-energy tensor in
state $\psi_0$ and certain other smooth local geometric terms, this
defect can be remedied. (The bound is then typically not a `closed
form' expression.)

\subsection{Macroscopic Stability: Passivity}

Pusz and Woronowicz introduced the notion of passivity in the following
way~\cite{PW}. Let $(\Af,\alpha_t)$ be a $C^*$-dynamical
system; that is, $\Af$ is a $C^*$-algebra, which we think of as the
algebra of observables for some quantum system, while $\alpha_t$ is the
map of evolution through time $t\in\RR$ corresponding to the undisturbed
evolution of the system, and has the group property
$\alpha_t\circ\alpha_{t'}=\alpha_{t+t'}$. 
Provided $\alpha_t$ is strongly continuous 
(i.e., the map $\RR\owns t\mapsto \alpha_t(A)\in\Af$ is continuous
for each $A\in\Af$)
we may define the generator $\delta$ of the evolution by
\begin{equation*}
\delta(A)=\left.\frac{\D}{\D t}\alpha_t(A)\right|_{t=0}
\end{equation*}
for the space of $A$ for which the derivative exists (in $\Af$), which we denote
$D(\delta)$. For example, if $\Af$ is the algebra of bounded operators on the Hilbert
space of a quantum mechanical system with Hamiltonian $H$, then
$\delta(A)=i[H,A]$. We also have  $\delta(A)= \alpha_t^{-1}\left(\D/\D
t\,\alpha_t(A)\right)$ for any $t$. The motivating idea of~\cite{PW} is to understand
thermodynamic stability of the dynamical system with respect to cyclical
changes of external conditions. One might think of a box of gas which is
compressed and then allowed to return to its initial volume. In the
current setting, a cyclical process occuring during time interval
$[0,T]$ may be modelled by a perturbed time evolution $\beta_t$ 
satisfying
\begin{equation*}
\beta_t^{-1}\left(\frac{\D}{\D t}\beta_t(A)\right) = \delta(A) + i[h_t,A]\;,
\end{equation*}
and $\beta_0={\rm id}$  where $t\mapsto h_t$ is a differentiable assignment of a self-adjoint
element $h_t\in\Af$ to each time $t$, and $h_t=0$ for $t\notin[0,T]$.

Suppose the system is initially in state $\omega$. Then the work
performed by the external agent driving the cyclical process is
\begin{equation*}
W_h = \int_0^T \D t\, \omega(\beta_t(\dot{h}_t)) \;,
\end{equation*}
and the state $\omega$ is said to be \emph{passive} if $W_h\ge 0$ for any
$h_t$, i.e., if no cyclical process can extract energy from the system.
Thus passivity isolates the property 
characteristic of the second law of thermodynamics in
Kelvin's formulation, where we think of the system as a thermal reservoir from which
we attempt to extract work. 

Pusz and Woronowicz proved
\begin{theorem} A state $\omega$ is passive if and only if
\begin{equation*}
i^{-1}\omega(U^*\delta(U)) \ge 0
\end{equation*}
for all $ U\in\Uf_1(\delta):= \Uf_1(\Af)\cap D(\delta)$, where 
$\Uf_1(\Af)$ is the identity-connected component of the unitary elements of $\Af$.
\end{theorem}

Particular examples of passive states are provided by ground and KMS
states, or mixtures thereof. A key feature of passivity is that it
introduces a definite thermodynamic `arrow of time'.

\subsection{From Mesoscopic to Macroscopic}

Let us now see how passivity may be obtained from QEIs, giving a
simplified and slightly modified version of the discussion
in~\cite{FVpassive}. We begin by introducing an abstract formulation of QEIs for 
$C^*$-dynamical systems, to which end we must first provide a notion of
the energy density. Accordingly, we assume that $\Af$ is the algebra of
observables of a system in a spacetime of the form
$\RR\times\Sigma$, for $\Sigma$ compact and Riemannian, with volume
measure $\D\mu(\ux)$. The evolution
$\alpha_t$ corresponds to time-translations on spacetime and is assumed
to be strongly continuous with generator $\delta$. 

As one would not expect the energy density to exist for all states, we
must specify a smaller class of states and a class of unitary elements
large enough to be dense in $\Uf_1(\delta)$, in a suitable sense, but which
preserves the state space. Accordingly, let 
$\Of$ be a $*$-subalgebra of $\Af$ with $\II\in\Of\subset
\bigcap_n D(\delta^n)$, and 
is large enough that any element of $\Uf_\II(\delta)$ may be
approximated arbitrarily well by unitary elements of $\Of$ with respect
to the graph norm of $\delta$. That is, to any $U\in\Uf_\II(\delta)$ there
is a sequence of unitaries $U_n\in\Of$ with $U_n\to U$ and
$\delta(U_n)\to\delta(U)$. In addition, let $\Sf$ be a convex set of
states of $\Af$ which is closed under operations in
$\Of$.\footnote{That is, for any $0\not=A\in\Of$ and $\omega\in\Sf$, 
we have $\omega(A^*A)>0$ and $\omega^A(B)=\omega(A^*BA)/\omega(A^*A)$ defines a state
$\omega^A\in\Sf$.} 

The energy density $\varrho(t,\ux)$ is assumed to obey:
\begin{enumerate}
\item For each $A,B\in\Of$ and $\varphi\in\Sf$,
$\varphi(A\varrho(t,\ux)B)$ is a $C^1$ function on $\RR\times\Sigma$.
\item The energy density generates the dynamics, and energy is
conserved, i.e,
\begin{equation}
\int_\Sigma \D\mu(\ux) \varphi(A[\varrho(t,\ux),B]C) =
\frac{1}{i}\varphi(A\delta(B)C)
\end{equation}
for arbitrary $A,B,C\in\Of$, $\varphi\in\Sf$ and $t\in\RR$. 
\end{enumerate}
Here, expressions of the form $\varphi(A\varrho(t,\ux)B)$ should be
taken as a convenient shorthand: what is more precisely meant is the
following. Let $\Ff$ be the subspace of continuous linear functionals
on $\Af$ generated by functionals of the form $C\mapsto {}_A\varphi_B(C):=\varphi(ACB)$ (for
$A,B\in\Of$, $\varphi\in\Sf$). Then the energy density is a linear map 
$\boldsymbol{\varrho}:\Ff\to C^1(\RR\times\Sigma)$, and our shorthand
notation $\varphi(A\varrho(t,\ux)B)$ means
$(\boldsymbol{\varrho}({}_A\varphi_B))(t,\ux)$. 

We are now in a position to define a general type of QWEI in this
setting, by analogy with the result~\eqref{eq:FEQI}. Our definition
differs slightly from that given in~\cite{FVpassive}.

\begin{definition} Let $\WW$ be a class of nonnegative integrable
functions of compact support on $\RR$. The system $(\cA,\alpha_t,\Of,\Sf,\varrho)$ obeys a static
QWEI (SQWEI) with respect to $\WW$ if, for some 
$\omega\in\Sf$, there exists a map $q_\omega:\WW\to L^1(\Sigma)$ such that
\begin{equation}
\int \D t\, f(t)\varphi({:}\varrho(t,\ux){:})\ge -q_\omega(f)(\ux) \qquad
\hbox{$\mu$-a.e. in $\ux$}
\label{eq:SQWEI}
\end{equation}
for all $\varphi\in\Sf$, 
where ${:}\varrho{:}=\varrho-\omega(\varrho)\II$. (In this case, the same
is true for all $\omega'\in\Sf$, as we may take
$q_{\omega'}(f)(\ux)=q_{\omega}(f)(\ux)+\int \D t\,
f(t)\omega'({:}\varrho(t,\ux){:})$.)
\end{definition}

We now state and prove one of the main results of~\cite{FVpassive}. 

\begin{theorem} If $(\cA,\alpha_t,\Of,\Sf,\varrho)$ obeys a SQWEI then $(\cA,\alpha_t)$ admits
at least one passive state.
\end{theorem}
\begin{proof}
Fix a reference state $\omega\in\Sf$ and choose $f\in\WW$ with $\int \D t\, f(t)
=1$ (we may assume $\WW$ is conic without loss). For unitary $U\in\Of$,
\begin{eqnarray}
\frac{1}{i}\omega(U^*\delta(U)) &=& \int_\Sigma \D\mu(\ux)\,
\omega(U^*[\varrho(t,\ux),U])\nonumber\\
&=& \int_\Sigma \D t\, f(t) \int \D\mu(\ux)\, \omega(U^*[\varrho(t,\ux),U])\nonumber\\
&=& \int_\Sigma \D\mu(\ux) \int \D t\, f(t) \omega(U^*{:}\varrho(t,\ux){:}U) \nonumber\\
&\ge & - \int_\Sigma \D\mu(\ux) \,q_\omega(f)(\ux) \;,
\end{eqnarray}
where we apply~\eqref{eq:SQWEI} with $\varphi$ defined by $\varphi(A)=\omega(U^*AU)$.
Because unitary elements of $\Of$ provide arbitrarily good
approximations to elements of $\Uf_1(\delta)$ we may choose unitaries
$U_n\in\Of$  such that
\begin{equation}
\frac{1}{i}\omega(U_n^*\delta(U_n))\longrightarrow c_\omega :=\inf_{U\in\Uf_1(\delta)}
\frac{1}{i}\omega(U^*\delta(U))\;,
\end{equation}
as $n\to\infty$, thereby deducing that 
\begin{equation}
c_\omega\ge -\int_\Sigma \D\mu(\ux)\, q_\omega(f)(\ux)>-\infty\;.
\end{equation}
If $c_\omega\ge 0$ then $\omega$ is passive and we are done, so
suppose instead that $c_\omega<0$. By the Banach--Alaoglu Theorem there exists a state $\omega^p$ on $\Af$ and a
subnet $U_{n(\sigma)}$ of the $U_n$ such that
\begin{equation}
\omega^p(A)= \lim_\sigma \omega(U_{n(\sigma)}^* A U_{n(\sigma)}) \qquad A\in\Af\;.
\end{equation}
To complete the proof, we calculate
\begin{eqnarray}
\frac{1}{i}\omega^p(U^*\delta(U)) &=& \lim_{\sigma}\frac{1}{i}\omega(U_{n(\sigma)}^*
U^*\delta(U)U_{n(\sigma)}) \nonumber\\
&=& \lim_\sigma \big[\underbrace{i^{-1}\omega((UU_{n(\sigma)})^*
\delta(UU_{n(\sigma)}))}_{\ge c_\omega} -
\underbrace{i^{-1}\omega(U_{n(\sigma)}^*\delta(U_{n(\sigma)}))}_{\to c_\omega} \big]\nonumber\\
&\ge & 0\;,
\end{eqnarray}
so $\omega^p$ is passive. \qed
\end{proof}

In~\cite{FVpassive} we also defined the notion of a state $\omega$ being
\emph{quiescent}, in terms of the behaviour of function
$q_\omega(f_\lambda)$ in the limit $\lambda\to 0^+$, where
$f_\lambda(t)=f(\lambda t)$. We showed that quiescent states are passive
(and even ground states, under additional clustering assumptions). 

Of course, we would like to see that this abstract set-up can be
realised in practice, and in particular, that it applies to quantum
field theory in static spacetimes with compact spatial section. 
Here, we encounter a problem with the scalar field because its
$C^*$-algebraic description in terms of the Weyl algebra with generators
$W(F)$ is not a $C^*$-dynamical system with respect to the time-translations
\begin{equation}
\alpha_t W(F) = W(F_t)\quad \hbox{where}\quad F_t(\tau,\ux)=F(\tau-t,\ux)\;.
\end{equation}
(This problem would not occur with the Dirac field, but less was known
about Dirac QEIs when~\cite{FVpassive} was written!) Instead one can generate
$\Af$ from objects of the form
\begin{equation}
\int \D t\, h(t) \alpha_t W(F) \qquad (h\in\CoinX{\RR})
\label{eq:form}
\end{equation}
formed in quasifree Hadamard Hilbert space representations of the Weyl
algebra; as shown in~\cite{FVpassive}, all the requirements of the abstract setting are fulfilled
with $\Sf$ equal to the set of finite convex combinations of Hadamard
states occuring as vectors in quasifree Hadamard representations of the
Weyl algebra. (Microlocal techniques turn out
to be exactly the right tools for this nontrivial check.) The $*$-algebra $\Of$ is
generated by operators of the form $\exp \I A$, where $A=A^*$ is a
polynomial in objects of the type~\eqref{eq:form}.

A further problem, however, is that the passive state obtained from the Banach--Alaoglu
theorem lives on
$\Af$, rather than the Weyl algebra itself. Given sufficient
regularity (e.g., energy compactness, believed to hold for this theory)
we may reconstruct a passive state on the Weyl algebra~\cite{FVpassive}.
Again, this problem would not arise for the Dirac field. 

\subsection{From Macroscopic to Microscopic}

Finally, we briefly discuss the last link in our circle of stability
conditions. In~\cite{SV1}, Sahlmann and Verch considered general topological
$*$-dynamical systems and defined a \emph{strictly passive} state to be a
mixture of ground and KMS states (at possibly different inverse
temperatures). Note that this is a stronger requirement than the usual
notion of passivity, as employed in~\cite{PW,FVpassive}. They also
introduced the notion of an \emph{asymptotic $n$-point correlation
spectrum} which generalises the wave-front set to this setting, and
formulated an appropriate generalisation of the microlocal spectrum
condition. When applied to linear quantum field theory on stationary
spacetimes, with respect to the stationary time evolution, the original
microlocal spectrum condition is recovered. 
They then proved
that strictly passive states obey the generalised microlocal spectrum
condition: the key ingredient in their argument is that both (strict) passivity
and the microlocal spectrum conditions share a common arrow of time.

\section{Connections with Nuclearity}

Quite recently, evidence has emerged to suggest the existence of a 
connection between QEIs and nuclearity criteria, with possibly
far-reaching implications. We will consider the original nuclearity
condition of Buchholz and Wichmann~\cite{BuchWich86} (for other closely
related criteria see, e.g.,~\cite{BuchPorr}). We work within the
algebraic approach to quantum field theory~\cite{Haag}, and consider a quantum
field theory described by a Hilbert space $\HH$, a strongly continuous unitary
representation 
$g\mapsto U(g)$ on $\HH$ of the universal cover of the proper orthochronous
Poincar\'e group $\widetilde{\PP}^\uparrow_+$, and a net of
von Neumann algebras $\cR(\OO)$, consisting of bounded operators on
$\HH$ and indexed by open bounded contractible spacetime regions $\OO$.
The following axioms are assumed to hold: \emph{isotony} ($\OO'\subset\OO$ implies
$\cR(\OO')\subset\cR(\OO)$); \emph{covariance} ($U(g)\cR(\OO)U(g)^{-1}=\cR(g\OO)$ for
$g\in\widetilde{\PP}^\uparrow_+$);  \emph{locality} ($\cR(\OO)$ and $\cR(\OO')$ commute
if $\OO$ and $\OO'$ are spacelike separated) and the \emph{spectrum condition}
(the generators of spacetime translations,
$P_\mu$, associated with the representation $U$, are self-adjoint
operators such that $P_0$ and $P_0^2-P_1^2-P_2^2-P_3^2$ are positive). 
Finally, we assume the existence of a unique vacuum state: 
namely, that the Hamiltonian $H=P_0$ has a simple eigenvalue at zero
with normalised eigenvector $\Omega$. 

Given any double cone $\OO_r$ based on a ball of radius $r$ and any
$\beta>0$, let 
\begin{equation}
\sN_{\beta,r}=\{\E^{-\beta H}W\Omega:W\in\cR(\OO_r)~{\rm s.t.}~W^*W=\II\}\;.
\end{equation}
This set may be regarded as the set of local vacuum excitations
associated with $\OO_r$, damped exponentially in the energy. 
The theory is said to obey the \emph{condition of nuclearity} if,
firstly, each $\sN_{\beta,r}$ is a \emph{nuclear} subset of $\HH$
[see below] and, secondly, there exist positive constants $c$, $n$,
$r_0$ and 
$\beta_0$ so that the corresponding \emph{nuclearity index}
$\nu(\sN_{\beta,r})$ obeys
\begin{equation}
\nu(\sN_{\beta,r}) \le \exp\left(c r^3\beta^{-n}\right)
\label{eq:nuc_bd}
\end{equation}
for all $0<\beta<\beta_0$ and $r>r_0$. This condition is therefore
a restriction on the number of local degrees of freedom available to the
theory. 

In the above, a subset $\LL$ of $\HH$ has nuclearity index $\nu(\LL)=\inf \Tr
|T|$, where the infimum is taken over the set of trace-class
operators $T$ so that $\LL$ is
contained within the image of the unit ball $\HH_{(1)}$ of $\HH$ under $T$, and
$\LL$ is said to be nuclear if it has a finite nuclearity
index.\footnote{By convention, an infimum over an empty set is infinite,
so this amounts to the assertion that there does exist a trace-class 
$T$ with $\LL\subset T\HH_{(1)}$.}

Despite its rather technical definition, the condition of nuclearity is
well-motivated from a physical viewpoint as the discussion
in~\cite{BuchWich86} makes plain: the nuclearity index can be
interpreted as a local partition function, and the form of the
nuclearity bound~(\ref{eq:nuc_bd}) is suggested by the requirement that
the associated pressure should remain finite in the thermodynamic limit and
scale polynomially with temperature (as is the case, for example, in the
Stefan--Boltzmann law). 

Buchholz and Wichmann verified in~\cite{BuchWich86} that the massive
free scalar field satisfies the condition of nuclearity, and remark that
the same is true of the system of countably many fields with masses
$m_j$ given suitable conditions on the density of states. Namely, the
sets $\sN_{\beta,r}$ are nuclear if~\cite{BuchWich86} and only
if~\cite{BuchJung86} $\sum_j \exp(-\beta m_j)<\infty$ for all sufficiently
small $\beta$; furthermore, the nuclearity index may be estimated from
above by
\begin{equation}
\nu(\sN_{\beta,r}) \le \exp\left(c\left(\frac{r}{\beta}\right)^3\sum_j\left|\log(1-\E^{-\beta
m_j/2})\right|\right)
\label{eq:nuc_est}
\end{equation}
for all sufficiently large $r$ and small $\beta$, and some constant $c$. 
It is convenient to introduce $N(u)$, the number of particle species
with mass below $u$ by
\begin{equation}
N(u) = \sum_j \vartheta(u-m_j)\;.
\end{equation}
The assumption that $N(u)$ grows polynomially, $N(u)=O(u^p)$ as
$u\to\infty$, is sufficient to show (using~\eqref{eq:nuc_est})
that~\eqref{eq:nuc_bd} is satisfied, for any $n>3+p$. 
It is tempting to conjecture that this condition is also necessary, but
this is currently an open question, and relies on finding better lower bounds on
the nuclearity index than are currently known. We will return to this
point below. 

We now present some circumstantial evidence for a connection between
nuclearity criteria and QEIs. Fix some inertial frame of reference in
Minkowski space and let $\varrho_j$ be the energy density of the
free field of mass $m_j$ with Hilbert space $\HH_j$ and vacuum state
$\Omega_j$. Let $\Had_j\subset\HH_j$ be the corresponding space of
Hadamard vector states. The Hilbert space of the full theory is the
tensor product
\begin{equation}
\HH = \bigotimes_j{}^{\Omega_j} \HH_j\;;
\end{equation}
that is, the completion with respect to the obvious inner product of the
set of finite linear combinations of product states $\bigotimes_j\xi_j$
in which all but finitely many of the $\xi_j$ are equal to $\Omega_j$.
We define the space of Hadamard states 
$\Had$ of the full theory to consist of finite linear
combinations of product states $\bigotimes_j\xi_j$ in which each
$\xi_j\in\Had_j$ and all but finitely many $\xi_j$ equal $\Omega_j$, and
then define the total energy density as follows: 
for any $\eta=\bigotimes_j\eta_j$ and $\xi=\bigotimes_j\xi_j$ in
$\Had$ we set
\begin{equation}
\ip{\eta}{\varrho(x)\,\xi} = \sum_j
\ip{\eta_j}{\varrho_j(x)\,\xi_j}\prod_{k\not=j}\ip{\eta_k}{\xi_k}
\end{equation}
(noting that only finitely many terms contribute to the sum, and that
each product involves only finitely many terms differing from unity) and then
extend by linearity to all $\eta,\xi\in\Had$. The left-hand side should
be regarded as a quadratic form on $\Had$, taking values in the space of smooth
functions on spacetime; clearly, any normal-ordered quantity could be
treated in this way, and no constraints on the $m_j$ have been imposed. 
Since the $j$'th component of the full theory obeys the
QWEI~\eqref{eq:FEQI} for each mass $m_j$,
\begin{equation}
\int \D t\,|g(t)|^2 \ip{\psi_j}{\varrho_j(t,0)\,\psi_j} \ge
-\frac{\|\psi_j\|_{\HH_j}^2}{16\pi^3} \int_0^\infty \D u\, |\widehat{g}(u)|^2 u^4 \vartheta(u-m_j)\;,
\end{equation}
for all Hadamard states $\psi_j\in\Had_j$,
the full theory obeys
\begin{equation}
\int \D t\,|g(t)|^2 \ip{\psi}{\varrho(t,0)\,\psi} \ge
-\frac{1}{16\pi^3}\int_0^\infty \D u\,|\widehat{g}(u)|^2 u^4 N(u)\;,
\label{eq:GFFQWEI}
\end{equation}
for any normalised $\psi\in\Had$. 
Accordingly, polynomial growth of $N$ is sufficient for the theory 
to admit a worldline QWEI with test-functions $g$ drawn from $\CoinX{\RR}$,
and it is possible to show that it is a necessary and sufficient condition if
certain scaling behaviour is required:\footnote{If $N(u)$ grows faster than
polynomially, one may still formulate QWEIs, but for weight
functions with sufficiently rapid decay in Fourier space. In particular,
this would generally exclude compactly supported weights.} 

\begin{theorem} Consider a generalised free field with discrete mass spectrum
described by $N(u)$. Let $p>0$. Then the following are equivalent:\\
1) $N(u)=O(u^p)$ as $u\to\infty$;\\
2) The generalised free field obeys the QWEI~\eqref{eq:GFFQWEI} for arbitrary
$g\in\CoinX{\RR}$, and the bound has asymptotic behaviour of 
order $O(\tau^{-(p+4)})$ as $\tau\to 0+$, if we replace $g$ by
$g_\tau(t)=\tau^{-1/2}g(t/\tau)$. 
\end{theorem}
The proof of this result will be reported elsewhere. An immediate
corollary is that the existence of a QWEI with polynomial scaling
implies that the Buchholz--Wichmann nuclearity
condition~\eqref{eq:nuc_bd} is satisfied for any $n>p+3$. 

All this raises two questions, which are being pursued in on-going work
with Porrmann and Ojima. First, can we show that~\eqref{eq:nuc_bd}
implies that $N(u)$ is polynomially bounded? If so, we would have an
equivalence between QWEIs and nuclearity for this model. This leads to
the second question: Can we
understand the link at a deeper level, or is it merely a coincidence,
with no more significance than that both are manifestations of the
uncertainty principle? 
A suitable understanding of this question might lead to a general
framework for establishing QEIs in general quantum field theories. 
Part of the problem is to identify the right question, of course, and it
may be that one or both of nuclearity or QEIs need to be carefully
(re)phrased or even replaced. These questions also require consideration
of lower bounds on nuclearity indices: here a potential stumbling block
is the technical definition of many of the quantities appearing in
discussions of nuclearity, which are therefore not easily amenable to
direct calculation even in the simplest cases. 
Indeed this provides pitfalls for the unwary, one of which we have
recently noted~\cite{FOP04}: in the mathematical literature there is a 
notion of $p$-nuclear map, whose definitions for $p>1$ and $p\le 1$ take
rather different forms. Although this difference has occasionally been
noted in the physics literature~\cite{Schumann}, one often finds the
$p\le 1$ definition used for all $p$. However, as we show in~\cite{FOP04}, the
corresponding nuclearity index would vanish identically for $p>1$ according to
this definition! Fortunately this confusion does not appear to have adverse
consequences in the literature so far, but it serves as a warning. 

\section{Conclusion}

Quantum Energy Inequalities are an expression of the uncertainty
principle, and as such are deeply rooted within quantum theory. It is
perhaps not surprising that they have connections with other fundamental
properties: unravelling
these interconnections has the potential to deepen our understanding of the
structure of quantum field theory and the nature of quantised matter.
Much remains to be done!
\bigskip

{\noindent\emph{Acknowledgment:} I am grateful to Lutz Osterbrink for a careful reading of the manuscript.}

%
%

%
%

\end{document}